\documentclass[twocolumn,preprintnumbers,amsmath,amssymb]{revtex4}
\usepackage{dcolumn}
\usepackage{bm}
\usepackage{graphicx}
\usepackage{amsmath}
\begin{document}
\title{Composite solitary vortices of three-wave mixing in quasi-phase-matched photonic crystals}
\author{Chao Kong$^{1,2}$ \ Jinqing Li$^{1,2}$ \ Xinyi Tang$^{1,2}$ \ Xuli Li$^{1,2}$ \ Ju Jiao$^{1,2}$ \ Jun Cao$^{1,2}$ \ Haiming Deng$^{1,2}$\footnote{Corresponding Author: woshidenghaiming@126.comn}}
\affiliation{$^1$ School of Physics and Electronic-Electrical Engineering,
Xiangnan University, Chenzhou 423000, China }
\affiliation{$^2$ Microelectronics and Optoelectronics Technology Key Laboratory of Hunan Higher Education, Xiangnan University, Chenzhou 423000, China }
\begin{abstract}

We report the composite vortex solitons of three-wave mixing propagate stably in a three-dimensional (3D) quasi-phase-matched photonic crystals (QPM-PhC).  The modulation of QPM-PhC is designed as a checkerboard pattern. The vortex solitons, composed by three waves ($\omega_{1,2,3}$) propagating through the lattices, exhibit a four-spotted discrete type, which gives rise to four distinct modes: zero-vorticity, vortex, anti-vortex, and quadrupole. The composite vortex solitons result from combinations of these modes and lead to four cases: vortex doubling, hidden vortices, vortex up-conversion, and anti-vortex up-conversion.  Our findings indicate that all solitons can propagate stably through the crystals for 10 centimeters; however, only the vortex-doubling case remains stable over longer distances. This work enhances the understanding of vortex beam manipulation within 3D QPM-PhCs.


\end{abstract}

\maketitle

\section{Introduction}

Stabilizing multi-dimensional solitons is a current focal point in the field of nonlinear optics \cite{solitonboobk}. In Kerr ($\chi^{(3)}$) nonlinear media, the self-focusing effect can result in the collapse or expansion of states when the power of the nonlinear beam exceeds or falls below a critical threshold in free space \cite{Fibich1999,Berge1998,2Fibich2002}. On the other hand, quadratic ($\chi^{(2)}$) nonlinear media offer a more favorable environment for the propagation of multi-dimensional fundamental optical solitons in free space \cite{1Buryaka2002,3Torruellas1995,4Torruellas2002}. However, for vortex solitons, the presence of strong azimuthal instability prevents both $\chi^{(2)}$ \cite{Minadrdi2001} and $\chi^{(3)}$ \cite{Malomedvortex} media from sustaining a stable vortex soliton in an unconfined setting. A potential solution to this issue involves introducing a competing higher-order nonlinearity to compensate the lower-order attractive nonlinearity \cite{5DeSalvo1992,6Bosshard1995,YSK,Buryak1995}. In this case, stable vortex solitons can be generated in competing quadratic-cubic \cite{7Towers2001,10Mihalache,8Trapani2000} and cubic-quintic \cite{Mihalache2000} nonlinear media. Similar phenomena have been observed in quantum droplets of Bose-Einstein condensates featuring competing mean-field (cubic) and beyond mean-field (quartic) effects \cite{Petrov,QDreview,YLi2018,YVK2018,Xiliang,Guilong,Guilong2024,Guihua,EALHenn,BLiu,YLi2017,LDong2022}. Nevertheless, the activation of higher-order nonlinear effects necessitates a significant increase in optical power, typically requiring beam intensities on the order of  $\sim10$ GW/cm$^2$ for the quadratic-cubic case, which often approaches the damage threshold of nonlinear crystals. Therefore, finding a method to stabilize vortex solitons in these nonlinear media without triggering competing higher-order effects remains a formidable challenge.

\begin{figure}[h!]
{\includegraphics[width=0.9\columnwidth]{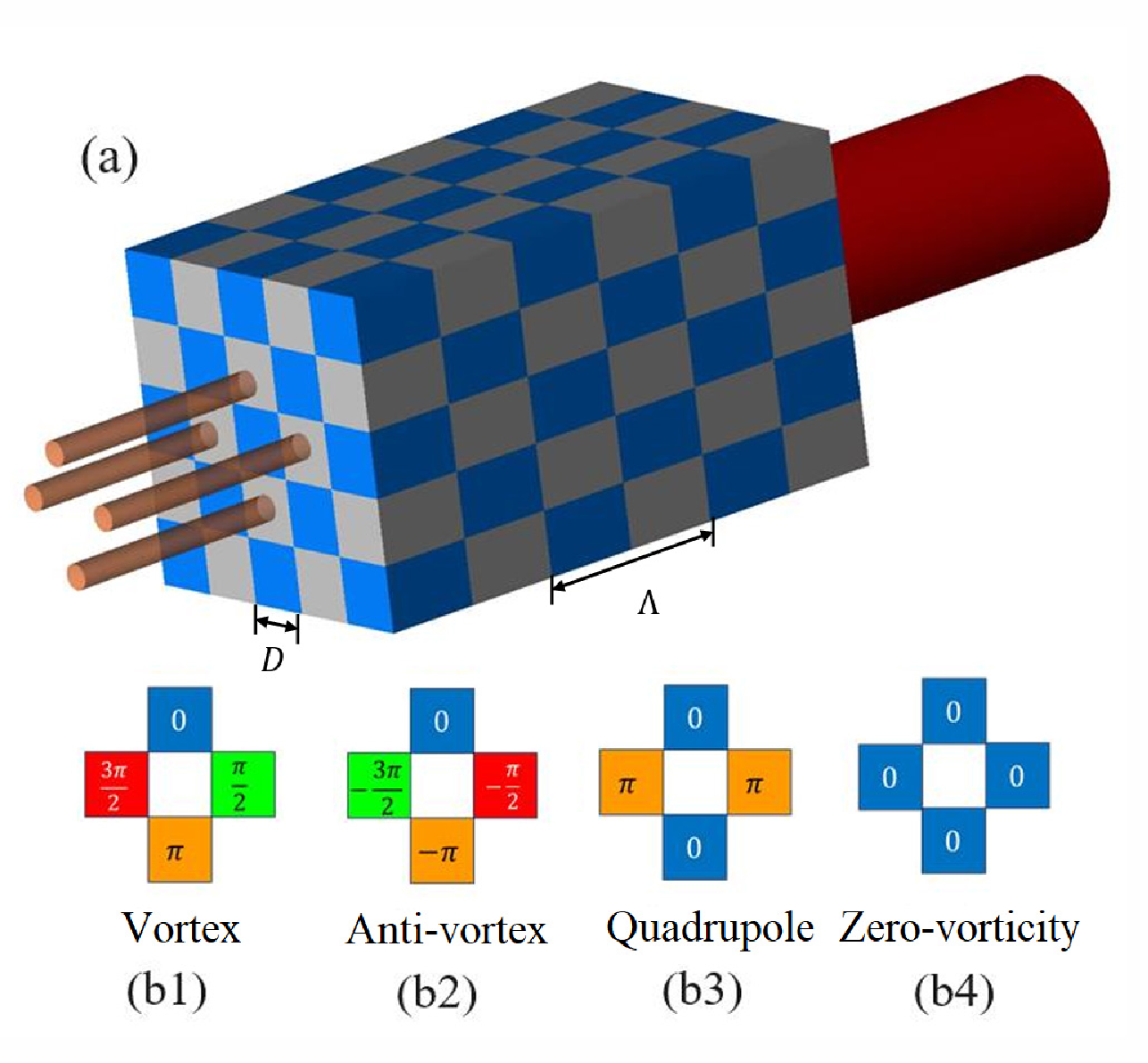}}
\caption{(a) The sketch for the 3D QPM-PhC and the propagation of the vortex solitons. $D$ is the length of the square cell, and $\Lambda$ is modulation period of QPM. (b1-b4) The phase structure for different modes (vortex, anti-vortex, quadrupole, and zero-vorticity) under the condition of four-spotted structure. }%
\label{crystal}%
\end{figure}

Recently, stable propagation of vortex solitons were predicted under the second-harmonic generation process in a three-dimensional (3D) quasi-phase-matched photonic crystal (QPM-PhC). The 3D QPM-PhC is characterized by a checkerboard structure \cite{Feiyan}. This structure is expected to be created through femtosecond laser processing on lithium niobate \cite{Xu2018,Wei2018,Tar2018,Review3D,9Arie,HLi2020,SLiu2023}. The vortex solitons emerge when a Laguerre-Gaussian laser beam of the fundamental frequency carrying an orbital angular momentum (OAM) of $l=1$ is injecting in the crystals. This type of soliton exhibits two distinct four-spotted structures: rhombus-shaped and square-shaped, each requiring different phase-matching conditions. The fundamental-frequency of the soliton carries a OAM with $l=1$, while the second-harmonic waves exhibit a quadrupole structure. By leveraging the QPM technique to spatially modulate the $\chi^{(2)}$ susceptibility of the crystal while maintaining a constant linear refractive index, this study paves the way for transmitting and manipulating solitary vortex bright solitons without a very strong power in purely $\chi^{(2)}$ media. The results could lead to a new way for the manipulation of light fields \cite{GLiu,GLiu2,AKarnieli,Ofir2022,Ofir2023,XZhang}.

The aim of this study is to investigate the propagation of vortex solitons under the conditions of three-wave mixing within the same type of 3D QPM-PhC. In contrast to the second-harmonic generation process described in Ref. \cite{Feiyan}, we introduce an additional wave into the system, providing the typical phase structure with the symmetry of the crystals. Our findings reveal that 4 types of composite vortex soliton be supported within the crystals by the combination of these typical phase structure. Their stabilities are verified through the direct simulation. This research significantly extends previous discoveries regarding stable vortex soliton propagation in nonlinear photonic crystals, offering a more practical approach. Very recently, a similar configuration was considered by a one-dimensional QPM-PhC and study the polarization of the dipole solitons in the three-wave mixing process \cite{Yuxin,Hengsu}. The subsequent sections are organized as follows: Section II presents the model, Section III showcases numerical results and discussions, and Section IV provides the conclusion.

\section{Model}

The objective of the present work is to achieve stable composite vortex solitons under the three-wave mixing process in 3D QPM-PhC with a checkerboard structure \cite{Driben, Yongyao2011, Warambhe}. The sketch map for this system is shown in Fig. \ref{crystal}(a). The paraxial propagation of the three light waves with frequencies satisfying $\omega_{3}=\omega_{1}+\omega_{2}$ can be governed by the coupling equations with the slowly varying amplitude approximation:
\begin{align}
&i\partial_{Z} A_{1}=-\frac{1}{2 k_1}\nabla^{2}A_{1}-\frac{2 d(Z,X,Y)\omega_1}{c n_1} e^{-i\Delta k z} A_{2}^{\ast}A_{3}, \label{A1}\\
&i\partial_{Z} A_{2}=-\frac{1}{2 k_2}\nabla^{2}A_{2}-\frac{2 d(Z,X,Y)\omega_2}{c n_2} e^{-i\Delta k z} A_{1}^{\ast}A_{3}, \label{A2}\\
&i\partial_{Z} A_{3}=-\frac{1}{2 k_3}\nabla^{2}A_{3}-\frac{2 d(Z,X,Y)\omega_3}{c n_3} e^{i\Delta k z} A_{1}A_{2}, \label{A3}
\end{align}
where $\nabla^{2}=\partial_{XX}+\partial_{YY}$ is the Laplace operator, $A_{1,2,3}$, $\omega_{1,2,3}$ ($\omega_{3}=\omega_{1}+\omega_{2}$), $n_{1,2,3}$ and $k_{1,2,3}$ are the slowly varying envelopes, circular frequencies, refractive indexes and wave vectors of the three waves, respectively, and $\Delta k = k_1+k_2-k_3$ is the phase mismatch. $d(Z,X,Y)$ is the spatially varying magnitude of the second-order susceptibility, which can be expressed by
\begin{eqnarray}
d(Z,X,Y)=\sigma(X,Y)d(Z), \label{dXYZ}
\end{eqnarray}
with
\begin{eqnarray}
&&\sigma(X,Y)=-\mathrm{sgn}\left[\cos(\pi X/D)\cos(\pi Y/D)\right], \label{sigma}\\
&&d(Z)=d_0\mathrm{sgn}\cos(2\pi Z/\Lambda), \label{dZ}
\end{eqnarray}
where $D$ is the length of each square cell of the checkerboard structure in the $(X,Y)$ plane, $d_0$ corresponds to the coefficient $\chi^{(2)}$, and $\Lambda$ is the period of the QPM modulation [see in Fig. \ref{crystal}(a)]. Eq. (\ref{dZ}) can be expanded by the Fourier transformation as \cite{Aviv2018,Aviv2022,Tal}
\begin{equation}
d(Z)=d_{0}\sum_{m\neq 0}\frac{2}{\pi m}{\sin }\left( \frac{\pi m}{2}\right) {%
\exp \left( i\frac{2\pi mZ}{\Lambda }\right) .}  \label{QPM}
\end{equation}
Here, in Eq. (\ref{QPM}), we have assumed that the duty cycle for the modulation of the QPM to be $1/2$. The adopted configuration of $d(X,Y,Z)$ in Eqs. (\ref{sigma},\ref{dZ}) defines a 3D QPM-PhC [see in Fig. \ref{crystal}(a)]. Substituting Eqs. (\ref{sigma}) and (\ref{QPM}) into Eqs. (\ref{A1}-\ref{A3}), then applying the following rescaling definitions \cite{ZFY,YLiTWM,CRP,Luther,Jiantao}:
\begin{align}
&{{I}_{0}}=\left( \frac{{{n}_{1}}}{{{\omega }_{1}}}+\frac{{{n}_{2}}}{{{\omega }_{2}}}+\frac{{{n}_{3}}}{{{\omega }_{3}}}\right){{\left| {{A}_{0}} \right|}^{2}} \notag\\
& {\psi_{j}}={{(\frac{{{\omega }_{j}}}{{{n}_{j}}}{{I}_{0}})}^{-\frac{1}{2}}}{{A}_{j}}{{e}^{i(\Delta k-2\pi/\Lambda)Z}},\quad j=1,2,3 \notag\\
& z^{-1}_{d} =\frac{2{d}_{ij}}{\pi c}{{(\frac{\omega_{1}\omega_{2}\omega_{3}}{n_{1}n_{2}n_{3}}{{I}_{0}})}^{\frac{1}{2}}}, \notag\\
& z =Z/z_{d},\quad x=X\sqrt{{{k}_{1}}/z_{d} },\quad y=Y\sqrt{{{k}_{1}}/z_{d}}, \notag\\
& \Omega =z_{d}(\Delta k-2\pi/\Lambda), \label{normalization}
\end{align}
Eqs. (\ref{A1}-\ref{A3}) can be rewritten in normalized form as
\begin{align}
&i\partial _{z}\psi _{1}=-\frac{1}{2}\nabla _{x,y}^{2}\psi _{1}-\Omega \psi
_{1}-2 \sigma(x,y) \psi _{2}^{\ast }\psi _{3},  \label{psi1} \\
&i\partial _{z}\psi _{2}=-\frac{1}{2\eta_{21}}\nabla _{x,y}^{2}\psi _{2}-\Omega \psi
_{2}-2 \sigma(x,y) \psi _{1}^{\ast }\psi _{3},  \label{psi2} \\
&i\partial _{z}\psi _{3}=-\frac{1}{2\eta_{31}}\nabla _{x,y}^{2}\psi _{3}-\Omega \psi
_{3}-2\sigma(x,y)\psi _{1}\psi_{2}, \label{psi3}
\end{align}
where $\nabla _{x,y}^{2}=\partial _{x}^{2}+\partial _{y}^{2}$, $A_{0}$ is a typical value of electric intensity for the $\chi^{(2)}$ process, $\eta_{ij}=k_{i}/k_{j}$, $\Omega$ is an effective detuning. Here, we have assumed only the oscillation with $m=\pm1$ in Eq. (\ref{QPM}) is the closest to the phase mismatching, which is similar to the rotating wave approximation.

Equations (\ref{psi1}-\ref{psi3}) conserve two dynamical invariants, \textit{viz}.,
the total Hamiltonian and power (alias the Manley-Rowe invariant \cite{Gil}%
),
\begin{align}
&H=\iint {(\mathcal{H}_{P}+\mathcal{H}_{\Omega }+\mathcal{H}_{\chi ^{(2)}})}%
dxdy  \label{H} \\
&P=\iint (|\psi _{1}|^{2}+|\psi _{2}|^{2}+2|\psi_{3}|^{2})dxdy=P_{1}+P_{2}+P_{3},\label{Norm}
\end{align}%
where
\begin{align}
&\mathcal{H}_{P}=\frac{1}{2}|\nabla {{\psi }_{1}}|^{2}+\frac{1}{\eta_{21}}|\nabla {{\psi }_{2}}|^{2}+\frac{1}{\eta_{31}}%
|\nabla {{\psi }_{3}}|^{2}, \notag \\
&\mathcal{H}_{\Omega }=-\Omega (|\psi_{1}|^{2}+|{\psi _{2}|}^{2}+|\psi_{3}|^{2}), \notag\\
&\mathcal{H}_{\chi ^{(2)}}=-2\sigma(x,y)\left(\psi^{\ast} _{1}\psi^{\ast}_{2}\psi _{3}+\mathrm{c.c}\right). \label{3Hams}
\end{align}

The estimation for the experimentally relevant characteristics of the system are carried out as follow: We adopt QPM-PhC fabricated with lithium niobate ($d_{0}=27$ pm/V) \cite{Suchowski}. For convenience, we assume the wavelength $\lambda_{1}=\lambda_{2}=1064$ nm with Type-II polarization, and $\lambda_{3}=532$ nm, and the refractive index for the three waves are $n_{1}\approx n_{2} \approx n_{3} = 2.2$. The amplitude of the electric field $A_{0}$ is selected as $A_{0}=200$ kV/cm \cite{Feiyan}. This assumptions yield that $z_{d}=0.0625$ cm. Hence, the units of dimensions, detuning, the intensities of wave function and the total power of the three waves are generated as follow: $x=1$ and $y=1$ correspond to 0.7 $\mu$m, $z=1$ corresponds to 0.0625 cm, $\Omega=1$ correspond to $16$ cm$^{-1}$, $|\psi_{1}|^2=1$ and $|\psi_{2}|^{2}=1$ correspond to 80 MV/cm$^{2}$, $|\psi_{3}|=1$ corresponds to 160 MV/cm$^{2}$, and $P=1$ corresponds to 40 W.

\section{Numerical results and discussions}
As mentioned above, we have assumed that the polarization of $\omega_1$ and $\omega_2$ are in the Type-II polarization and satisfying $\omega_1=\omega_2=\omega$, hence, $\omega_{3}=2\omega$, which yields that $\eta_{21}=1$ and $\eta_{31}=2$. The bright solitons of Eqs. (\ref{psi1}-\ref{psi3}) can be represent by
\begin{eqnarray}
\psi_{1,2,3}(x,y,z)=\phi_{1,2,3}(x,y)\exp\left(i\beta_{1,2,3}z\right). \label{soliton}
\end{eqnarray}
where $\phi_{1,2,3}$ are the stationary solution of the three waves with $\beta_{1,2,3}$ being the propagation constants of them.

According to the phase matching condition of the three-wave mixing in sum-frequency generation, the propagation constants and the phase structure of all the components are satisfying
\begin{align}
&\beta_{3}=\beta_{1}+\beta_{2}, \label{beta}\\
&\mathrm{Arg}\left[\phi_{3}(x,y)\right]=\mathrm{Arg}\left[\phi_{1}(x,y)\right]+\mathrm{Arg}\left[\phi_{2}(x,y)\right].  \label{Phase-matching}
\end{align}
The vortex solitons are split into distinct four spots due to the instability and locked four cells with the same phase-matching condition, leading to a discrete form of square-shaped phase structures \cite{Disresoli,Zibin}. According to symmetries for a square which can be characterized by the dihedral D4 group, the maximum absolute value of topological charge is 1 \cite{Ferrando2005}. Positive and negative values correspond to vortex and anti-vortex phase structures, respectively [see in Fig. \ref{crystal}(b1,b2)]. Doubling these two vortex phase structures produces a quadrupole structure with a $\pi$-phase shift between adjacent spots [see in Fig. \ref{crystal}(b3)]. Finally, if the four peaks of a single mode share the same phase, the phase structure can be referred to as a zero-vorticity type [see in Fig. \ref{crystal}(b4)].

For convenience, we firstly fix $\Omega=0$, which assumes that the three waves are under the perfect phase-matching with the QPM-PhC, and fix $\omega_{1}$ contains a vortex structure. According to the request of the phase-matching condition, there are only 4 types of composite vortex modes supported by the system:
\begin{align}
& \mathrm{Vortex}+\mathrm{Vortex}=\mathrm{Quadrupole}, \label{VD}\\
& \mathrm{Vortex}+\mathrm{Anti\text{-}vortex}=\mathrm{Zero\text{-}vorticity}, \label{HV}\\
& \mathrm{Vortex}+\mathrm{Zero\text{-}vorticity}=\mathrm{Vortex}, \label{VU}\\
& \mathrm{Vortex}+\mathrm{Quadrupole}=\mathrm{Anti\text{-}vortex} \label{AVU}
\end{align}

\begin{figure}[h!]
{\includegraphics[height=0.55\columnwidth]{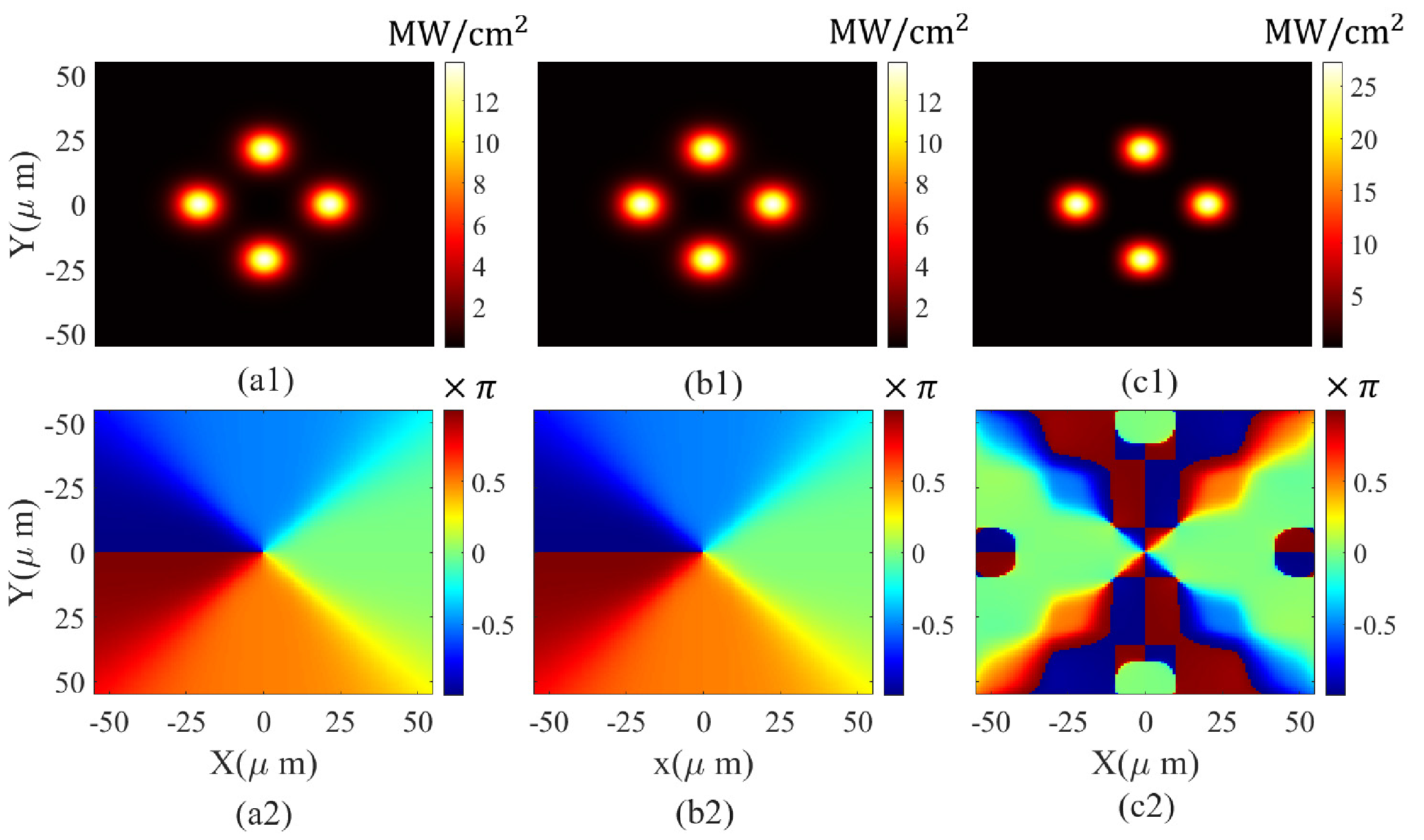}}
\caption{Example of the composite vortex soliton with $1+1=2$. (a1,b1,c1) and (a2,b2,c2) Intensity patterns and the vortex structure of the 3 components, respectively. Here, we select $P=1.2$ kW and $D=2.1$ $\mu$m.}\label{mode1}%
\end{figure}
\begin{figure}[h!]
{\includegraphics[height=0.55\columnwidth]{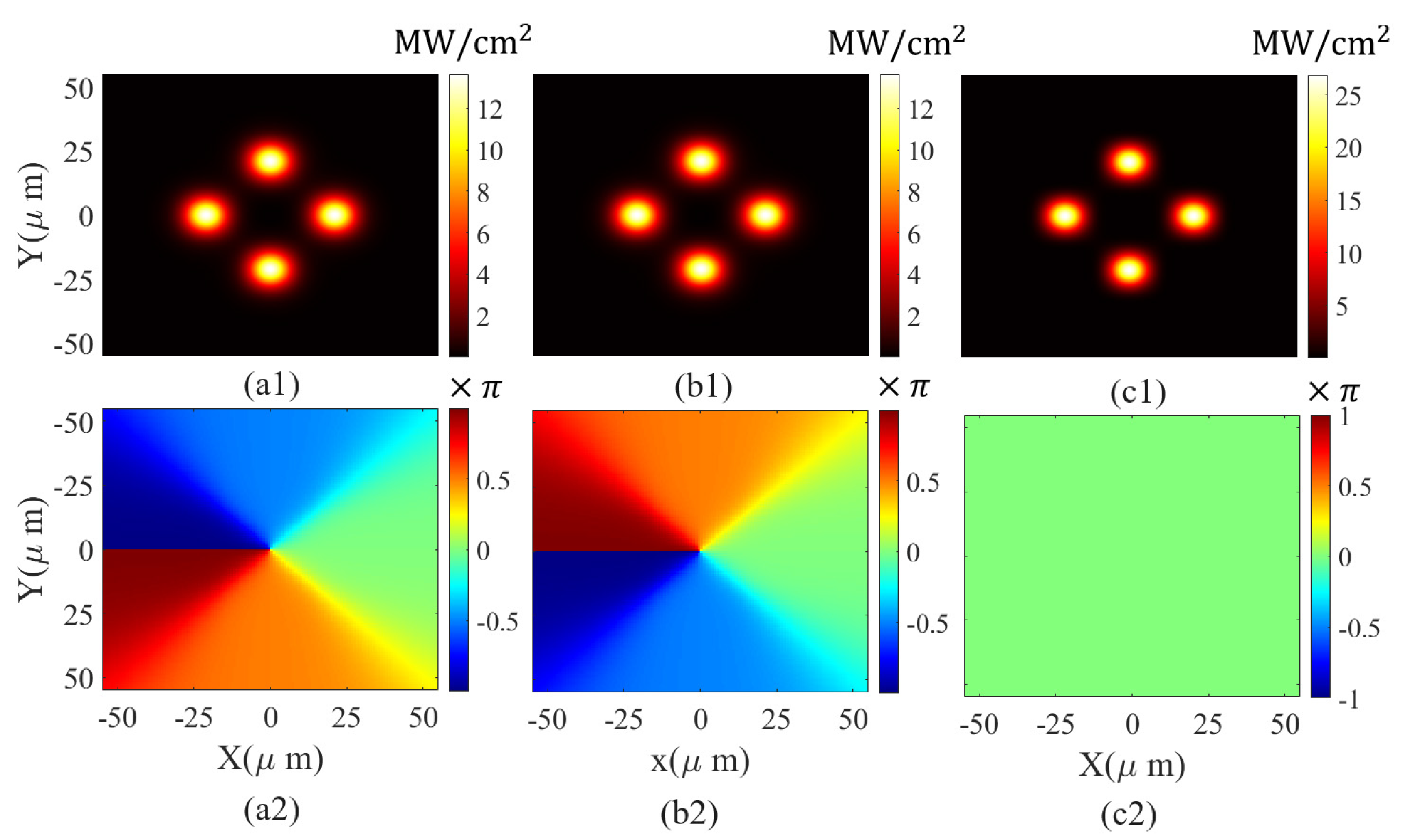}}
\caption{Example of the composite vortex soliton with $1-1=0$. (a1,b1,c1) and (a2,b2,c2) Intensity patterns and the vortex structure of the 3 components, respectively. Here, we select $P=1.2$ kW and $D=2.1$ $\mu$m. } \label{mode2}%
\end{figure}
\begin{figure}[h!]
{\includegraphics[height=0.55\columnwidth]{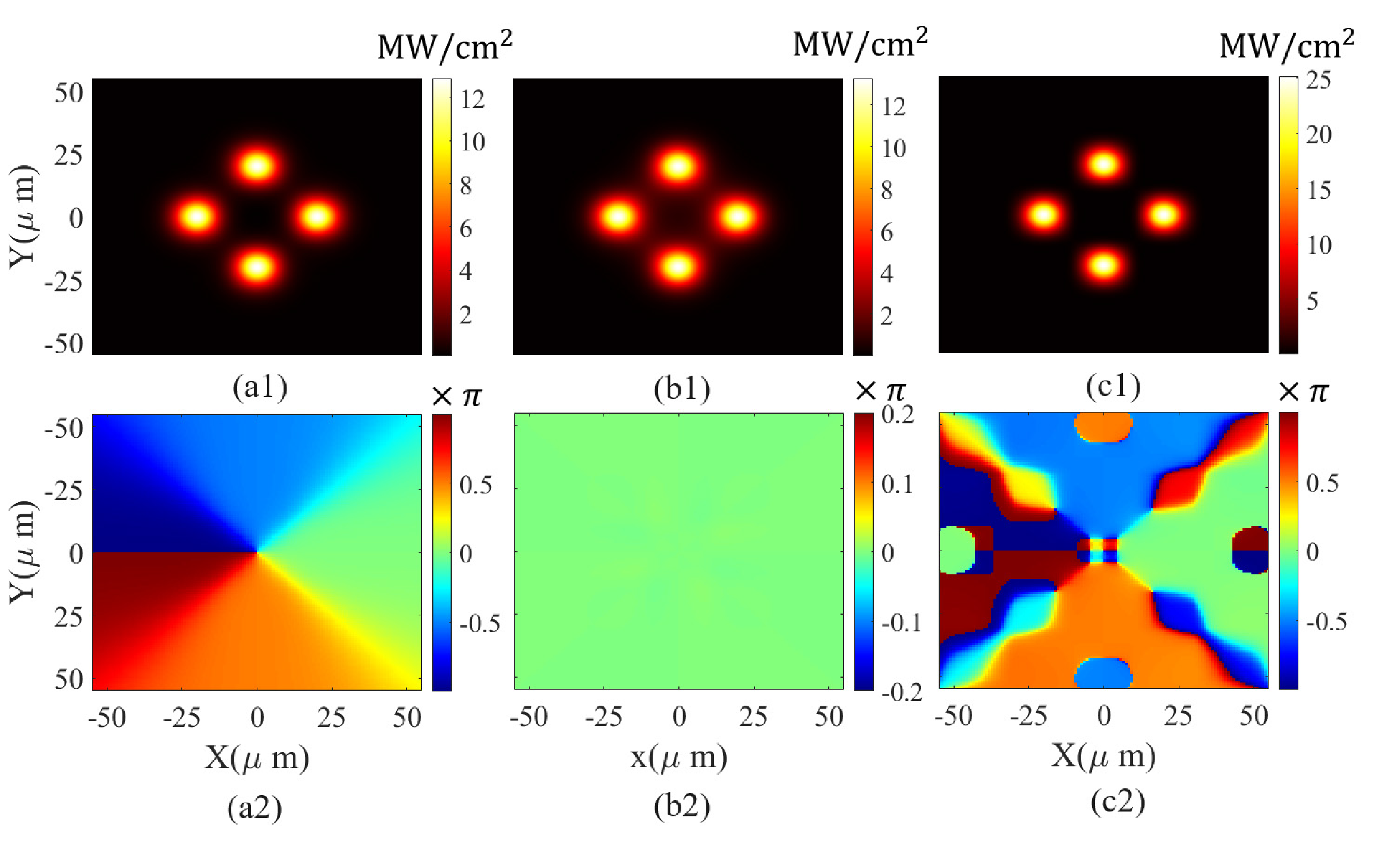}}
\caption{Example of the composite vortex soliton with $1+0=1$. (a1,b1,c1) and (a2,b2,c2) Intensity patterns and the vortex structure of the 3 components, respectively. Here, we select $P=1.2$ kW and $D=2.1$ $\mu$m. } \label{mode3}%
\end{figure}
\begin{figure}[h!]
{\includegraphics[height=0.55\columnwidth]{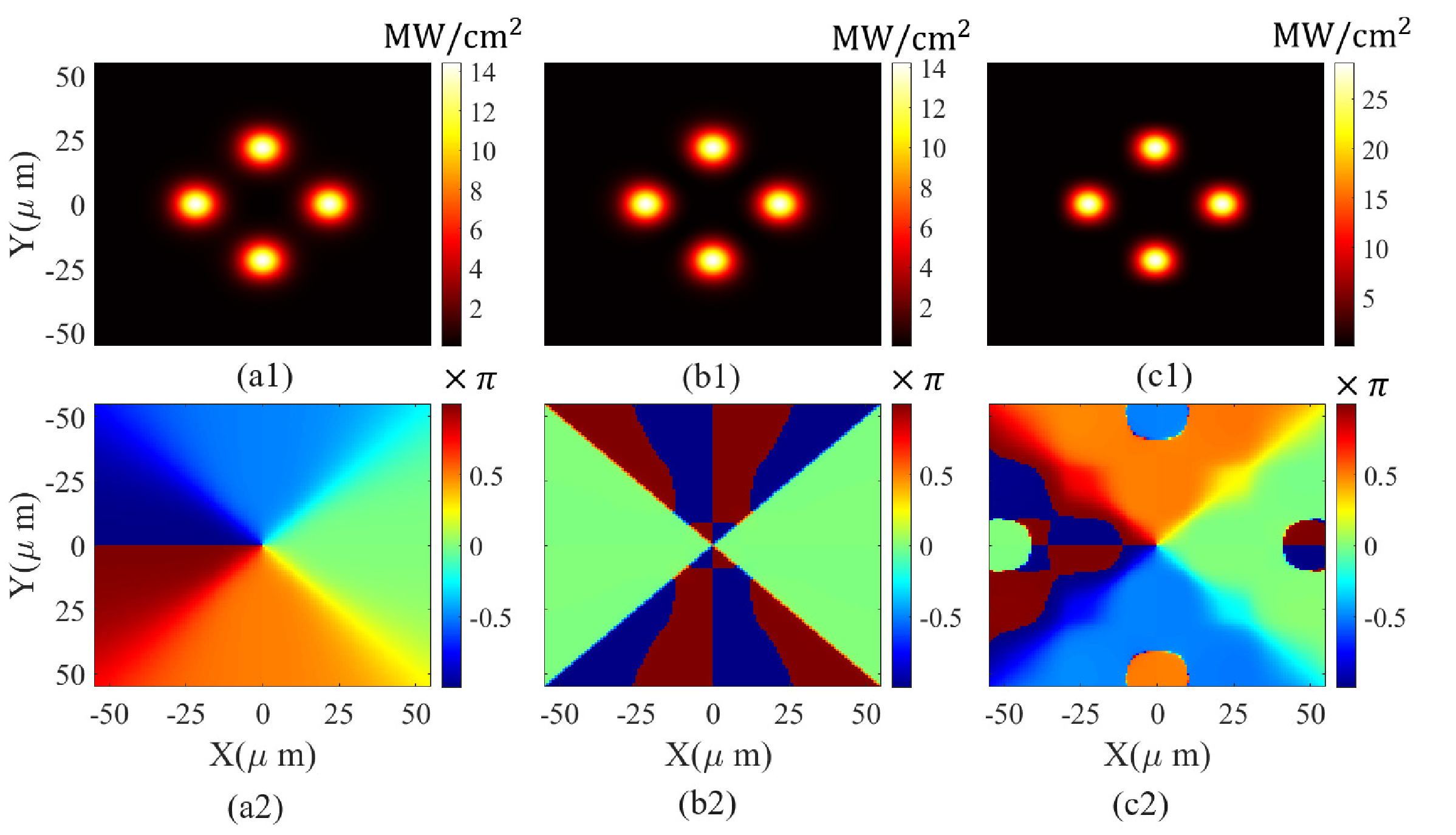}}
\caption{Example of the composite vortex soliton with $1-2=-1$. (a1,b1,c1) and (a2,b2,c2) Intensity patterns and the vortex structure of the 3 components, respectively. Here, we select $P=1.2$ kW and $D=2.1$ $\mu$m. }\label{mode4}%
\end{figure}

In Eq. \ref{VD}, the quadrupole mode in $\omega_{3}$ can be view as a doubling of the vortex mode from $\omega_{1,2}$; therefore, we can refer to this phenomenon as `vortex doubling'. In Eq. (\ref{HV}), the phase structure of the $\omega_{1}$ (vortex) and the $\omega_{2}$ (anti-vortex) cancels out each other, resulting in the conversion to zero-vorticity structure in $\omega_{3}$. We designate this case as `hidden vorticity'. In Eqs. (\ref{VU}, \ref{AVU}), both the vortex and anti-vortex structure are convert into $\omega_{3}$ through a zero-vorticity and a quadrupole structure in $\omega_{2}$, respectively; we refer to these scenarios respectively as `vortex up-conversion' and `anti-vortex up-conversion'. Additionally, since the quadrupole structure can be considered a doubling either the vortex or anti-vortex structure, we can represent it using the numbers $2$ or $-2$. The zero-vorticity can be represented by the number $0$. Hence, Eqs. (\ref{VD}-\ref{AVU}) can be represented by 4 numeric expressions as
\begin{eqnarray}
&& 1+1=2, \quad\mathrm{Vortex\thinspace doubling} \notag\\
&& 1-1=0, \quad\mathrm{Hidden\thinspace vortices} \notag\\
&& 1+0=1, \quad\mathrm{Vortex\thinspace up\text{-}conversion} \notag\\
&& 1-2=-1 \quad\mathrm{Anti\text{-}vortex\thinspace up\text{-}conversion} \notag
\end{eqnarray}

Typical examples of the composite vortex solitons of these modes, obtained using the imaginary-time method \cite{ITM}, are displayed in Figs. \ref{mode1}-\ref{mode4}, the control parameters for the solutions are $P$ and $D$. Numerical simulation finds that there is a threshold power, namely $P_{\mathrm{cr}}$, for supporting these soliton solutions. When $P<P_{\mathrm{cr}}$, the imaginary-time method cannot converge to a vortex solution. Only fundamental solitons can be produced under this circumstance. The value of $P_{\mathrm{cr}}$ with $\Omega=0$ for each composite vortex modes are shown in TABLE I.

\begin{table}
\caption{\label{tab:table1} $P_{\mathrm{cr}}$ for each composite vortex modes at $\Omega=0$.}
\begin{ruledtabular}
\begin{tabular}{cccccccccccccccccccccc}
&Vortex doubling & $320 W$\\
\ \ \ \ \ &Hidden vortices&$280 W$ \\
\ \ \ \ \ &Vortex up-conversion&$200 W$ \\
\ \ \ \ \ &Anti-vortex up-conversion&$400 W$\\
\hline
\end{tabular}
\end{ruledtabular}
\end{table}

\begin{figure}[h!]
{\includegraphics[height=0.55\columnwidth]{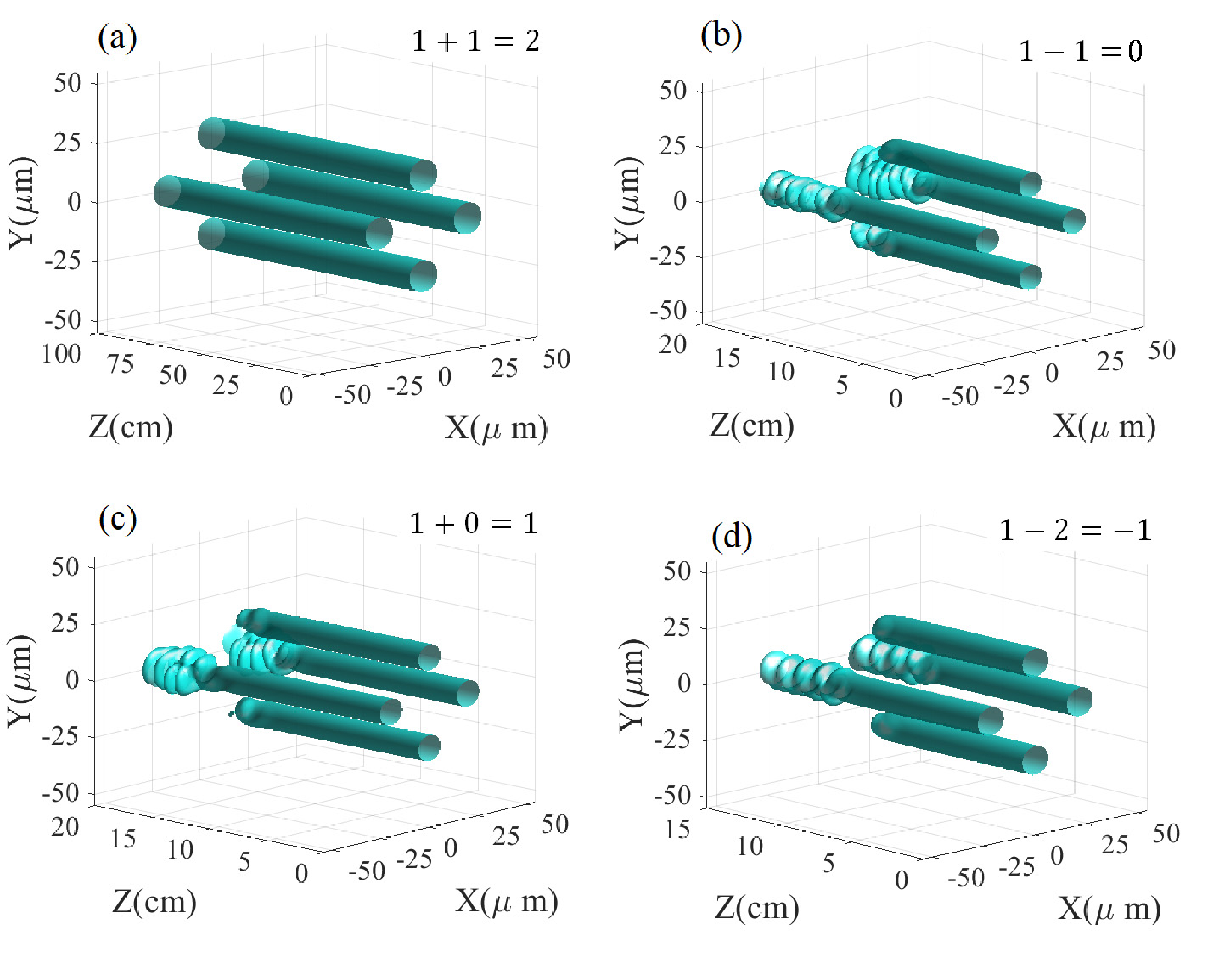}}
\caption{Direct simulations to the composite vortex solutions in Figs. \ref{mode1}-\ref{mode4}, which is shown by the isosurface of the effective total intensity pattern $|\psi_1(x,y)|^2+|\psi_{2}(x,y)|^2+2|\psi_{3}(x,y)|^2$. Here, only the composite vortex mode of `Vortex doubling' can survive at least up to 100 cm, the `Hidden vortices' and `Vortex up-conversion' modes are stable for approximately $Z=16$ cm, while the `Anti-vortex up-conversion' mode is stable for around $Z=11$ cm.}\label{Realtime}%
\end{figure}

The stability of these composite vortex solitons is confirmed by direct simulations in Eqs. (\ref{psi1}-\ref{psi3}) with 1\% noises added to the solution. Numerical simulations show that these composite vortex soliton solutions maintain their intensity pattern unchanged for at least up to $Z=10$ cm. This length is significantly longer than ten times the diffraction length of the soliton solutions and is also sufficient for the fabrication of the crystals, demonstrating the stability of these solitons. However, upon extending the propagation length, only the composite vortex mode of `Vortex doubling' can survive at least up to 100 cm, the `Hidden vortices' and `Vortex up-conversion' modes remain stable for approximately $Z=16$ cm, while the `Anti-vortex up-conversion' mode is stable for around $Z=11$ cm. Increasing the strength of the noise perturbations can weaken the stability of these three modes. As a result, only the "Vortex doubling" mode demonstrates robust stability, while the other three modes exhibit much weaker stability. Direct simulation for these 4 types of modes are illustrated in Fig. \ref{Realtime}.

\begin{figure}[h!]
{\includegraphics[height=0.55\columnwidth]{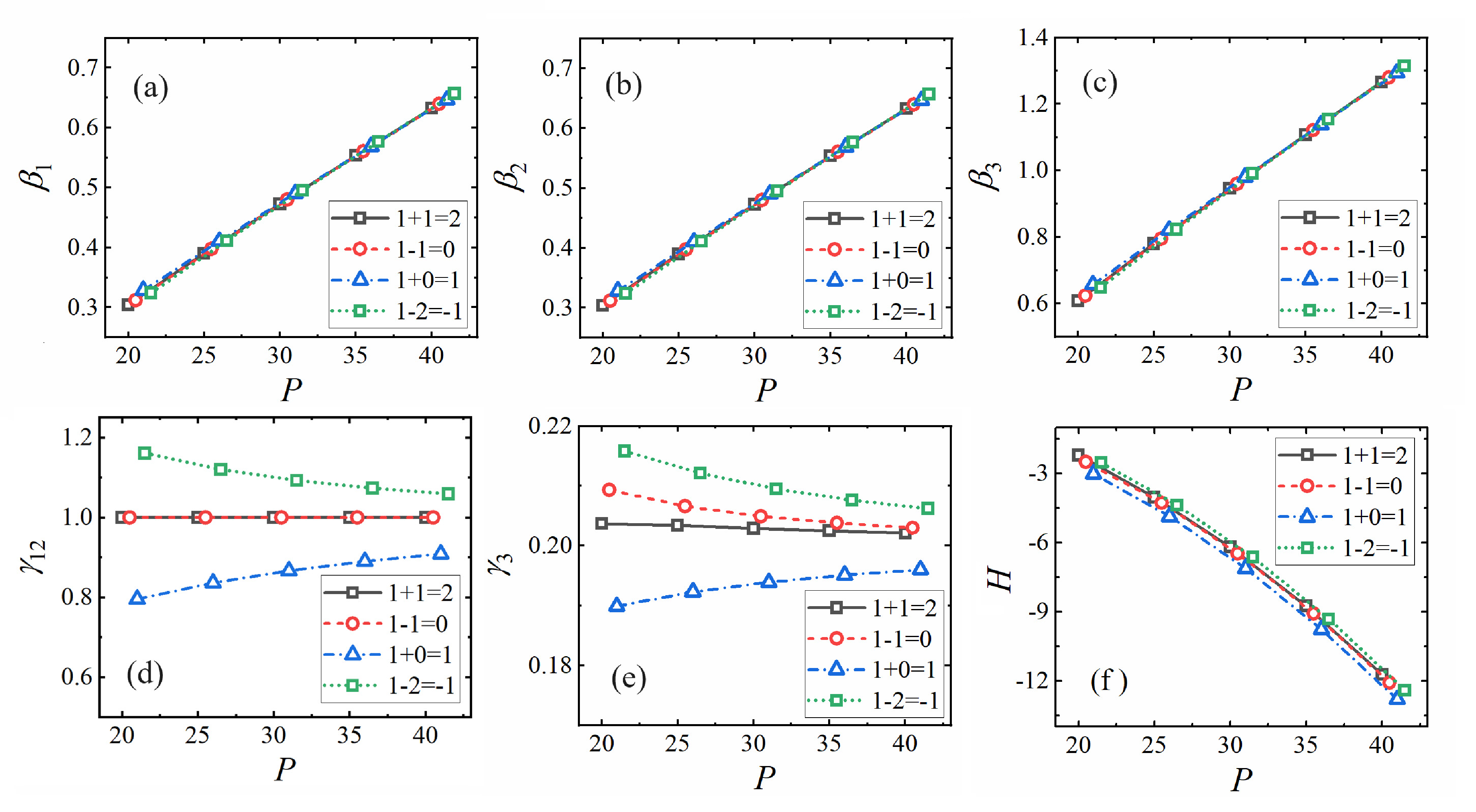}}
\caption{Propagation constants of the three waves, $\beta_{1,2,3}$, power sharing between the three waves, $\gamma_{1,2}$ and $\gamma_{3}$, and the Hamiltonian, $H$, of the 4 types of vortex solutions versus $P$. Here for all the solutions, we select $D=2.1$ $\mu$m. All the solutions in these panels are stable at least up to 10 cm. }\label{character}%
\end{figure}

To study the properties of the 4 types of composite vortex solitons, we introduce the power sharing between the 3 waves as $\gamma_{21}=P_1/P_2$ and $\gamma_{3}=P_{3}/P$. Therefore, various characteristics, including, propagation constants of the three waves, $\beta_{1,2,3}$, power sharing among the 3 waves, $\gamma_{21}$ and $\gamma_{3}$, and the Hamiltonian, $H$, of the 4 types of soliton solutions are presented as the function of $P$ in Fig. \ref{character}.

Figs. \ref{character}(a,b,c) illustrate that the relationships of the propagation constants follows Eq. (\ref{beta}) and satisfies $\beta_{1}=\beta_{2}=\beta_{3}/2$. The constraint relationship is necessitated by the frequencies relationships of $\omega_{1}+\omega_{2}=\omega_{3}$ and $\omega_{1}=\omega_{2}=\omega_{3}/2$. It shows that all the propagation constants are linear increase as the increase of $P$. Furthermore, $d\beta_{i}/dP>0$ ($i=1,2,3$) indicate that the soliton solutions adhere the Vakhitov-Kolokolov criterion \cite{VK}, a necessary condition for stable soliton supported by the focusing nonlinearity. In Fig. \ref{character}(d), it is observed that in the composite vortex modes with `Vortex doubling' and `Hidden vortices', the powers carried by $\omega_{1}$ and $\omega_{2}$ are identical ($P_1=P_2$). However, in the cases of `Vortex up-conversion' and `Anti-vortex up-conversion', the power relationship for $\omega_{1}$ and $\omega_{2}$ are $P_{1}>P_{2}$ and $P_{1}<P_{2}$, respectively. Notably, $\gamma_{21}$ for these two cases converges to 1 as $P$ increases. The power sharing of $\gamma_{3}$ is depicted in Fig. \ref{character}(e), showing that $\omega_{3}$ holds $\sim20$\% of the power in these 4 cases. Finally, in Fig. \ref{character}(f), the Hamiltonians of cases `Vortex doubling' and `Hidden vortices' are degenerated, case of `Vortex up-conversion' owns a lowest energy, while the case of `Anti-vortex up-conversion' exhibits the highest energy.

\begin{figure}[h!]
{\includegraphics[height=0.4\columnwidth]{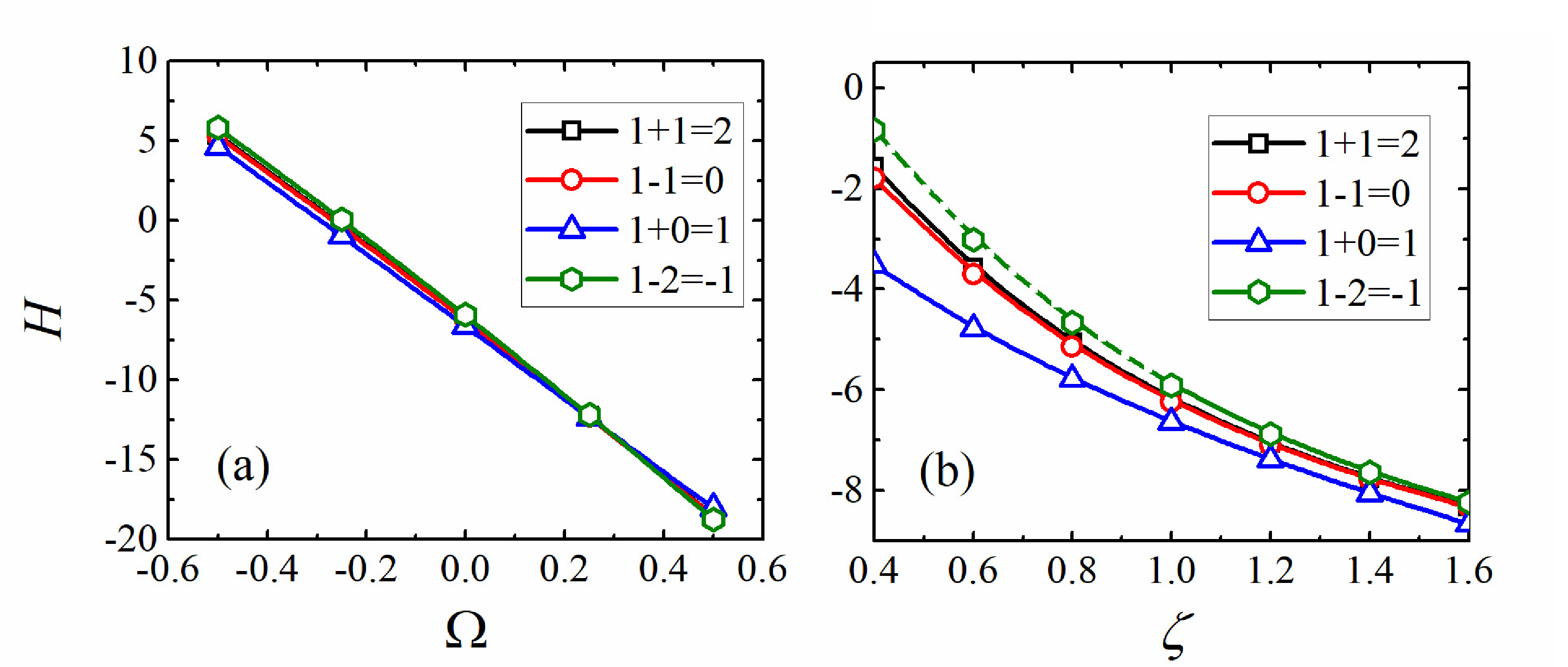}}
\caption{Hamiltonian of the vortex solitons versus $\Omega$ (a) and $\zeta$ (b). Here, the stability of the soliton is verified by length of 10 cm. Solid curve means soliton can survive at least up to 10 cm. The dashed curve means soliton cannot survive up to 10 cm. The composite vortex mode of `Anti-vortex up-conversion', which have the largest value of Hamiltonian among these 4 modes, destroy at $Z\approx9$ cm when $\omega_{2}<\omega_{1}$. In panel (a), we have selected $\omega_{1}=\omega_{2}$, while in panel (b), we select $\Omega=0$. For all the panels, we select $P=1.2$ kW and $D=2.1$ $\mu$m.}\label{Otherpro}%
\end{figure}

Finally, vortex soliton with the conditions of $\Omega\neq0$ and $\omega_{1}\neq\omega_{2}$ are also considered. For the case of $\Omega\neq0$, numerical simulations show that composite vortex solitons can be find in the finite range of $\Omega\neq0$. Fig. \ref{Otherpro}(a) displays 4 types of composite vortex solitons versus $\Omega$ at the range of $-0.5<\Omega<0.5$. It shows that the Hamiltonian almost linear decrease as the $\Omega$ increase. In the case of $\omega_{1}\neq\omega_{2}$, we can assume that $\omega_{2}=\zeta\omega_{1}$, then $\omega_{3}=(1+\zeta)\omega_{1}$. Neglecting the small difference between $n_{1}$, $n_{2}$ and $n_{3}$, one can determine that $\eta_{21}=\zeta$ and $\eta_{31}=(1+\zeta)$. Therefore, $\zeta$ is used to characterize the case of $\omega_{1}\neq\omega_{2}$. Fig. \ref{Otherpro}(b) displays the Hamiltonian versus $\zeta$. It shows that soliton solutions can still exist in the case $\omega_{1}\neq\omega_{2}$. The stability of the soliton in Fig. \ref{Otherpro} are verified by the direct simulation with the length of 10 cm. Solid curves indicate that the solitons can survive at least 10 cm, while the dashed curve indicates that the solitons cannot survive for up to 10 cm. The numerical simulation shows that the composite vortex mode of `Anti-vortex up-conversion', which still has the largest value of Hamiltonian among these 4 modes, is destroyed at $Z\approx9$ cm when $\omega_{2}<\omega_{1}$.

\section{conclusion}
In conclusion, we discuss the composite vortex solitons in a 3D quasi-phase-matched photonic crystal (QPM-PhC) with a checkerboard structure. The solitons are produced in the process of three-wave mixing processes through this crystal. Our studies reveal that four types of composite vortex solitons: vortex doubling (`$1+1=2$'), hidden vortices (`$1-1=0$'), vortex up-conversion (`$1+0=1$'), and anti-vortex up-conversion (`$1-2=-1$'), each characterized by the phase-matching condition of the three waves. Through numerical simulations, the study confirms the stability properties of these composite vortex solitons during propagation within the photonic crystals. Notably, the composite vortex  mode of `Vortex doubling' exhibits a highest stability, maintaining its intensity patterns unchanged at least up to $Z=100$ cm, indicating their robustness. Conversely, the other types of modes demonstrate stability only up to $10\sim15$ cm. Even though this length is already longer than ten times the diffraction length of the soliton solutions and is also sufficient for the fabrication of the crystals, their stability is much weaker than the `Vortex doubling' mode. The research extends previous findings on stable vortex soliton propagation in nonlinear photonic crystals, offering practical insights into the behavior of vortex solitons in such environments. By analyzing power sharing among the three waves and Hamiltonian values, the study offers a comprehensive understanding of the characteristics and dynamics of composite vortex solitons in the context of nonlinear optics. The results contribute to advancing knowledge in optical signal processing and communication systems, highlighting the potential applications of composite vortex solitons in these fields.

The analysis can be extended in other directions. Our current analysis focuses on the composite vortex soliton within a rhombus-shaped configuration, denoted by a phase-matching relationship of $\varphi_{3}=\varphi_{2}+\varphi_{1}$ (where $\varphi_{1,2,3}=\mathrm{Arg}\left[\phi_{1,2,3}(x,y)\right]$). While for the current checkerboard structure in the $(x,y)$ plane, there exists addition phase-matching relationship $\varphi_{3}=\varphi_{2}+\varphi_{1}-\pi$, which can give rise to square-shaped vortex solitons in 4 types of OAM modes. It may also relevant to cascade the cubic(Kerr) self-defocusing nonlinearity into the current system \cite{Xuxiaoxi}, and studies the optical vortex droplets states in 3D QPM-PhC with three-wave mixing process with different OAM relationship. Finally, the current discussion may also extend to the system of atomic-molecule Bose-Einstein condensates \cite{Alexander,Jieliu,Jieliu2}, to manipulate the conversion of vortices from atomic state to molecule states.

\section*{Acknowledgments}

The authors appreciate the useful discussions with Professor Yongyao Li and Professor Bin Liu (Foshan Univeristy). This work was supported by the Project of Hunan Provincial Education Office under Grants No. 23A0593, No. 23B0774, Hunan Provincial Natural Science Foundation of China under Grant No.2024JJ5364, Scientific Research Foundation of Xiangnan University for High-Level Talents, the Applied Characteristic Disciplines of Electronic Science and Technology of Xiangnan University (XNXY20221210), and Science and Technology Innovative Research Team in Higher Educational Institutions of Hunan Province.

\end{document}